# Intrinsic emittance properties of an Fe-doped β-$Ga_2O_3$(010) photocathode: Ultracold electron emission at 300 K and the polaron self-energy

Louis A. Angeloni[1], Ir-Jene Shan[1], J. H. Leach[2], and W. Andreas Schroeder[1]

[1] *Department of Physics, University of Illinois Chicago, 845 W. Taylor St., Chicago, IL 60607, USA*

[2] *Kyma Technologies Inc., 8829 Midway West Rd., Raleigh, NC 27617, USA*

**Abstract**

Measurements of the spectral emission properties of an iron-doped a β-$Ga_2O_3$(010) photocathode at 300 K reveal the presence of an ultracold contribution to the total electron beam emission with a 6 meV mean transverse energy (MTE) in the 3.5-4.4 eV photon energy range (282-354 nm). This extreme sub-thermal photoemission signal is consistent with direct emission of electrons photoexcited from the Fe dopant states into the low effective mass and positive electron affinity primary conduction band, and it is superimposed on a stronger signal with a larger MTE associated with an (optical)phonon-mediated momentum resonant Franck-Condon (FC) emission process from a thermally populated and negative electron affinity upper conduction band. For photon energies above 4.5 eV, a transition from a long to a short transport regime is forced by an absorption depth reduction to below 100 nm and both MTE signals exhibit spectral trends consistent with phonon-mediated FC emission if the polaron formation self-energy is included in the temperature of the initial thermalized photoexcited electron distribution.

Corresponding author: W. Andreas Schroeder, andreas@uic.edu

## Introduction

Understanding the fundamental physical processes involved in photoemission is key to the development (or discovery) of photocathode materials capable of delivering high brightness electron beams.[1-4] Of particular interest are laser-driven pulsed photoelectron sources with a low transverse beam divergence (i.e., initial mean transverse energy (MTE) of the photoemitted electrons) as this will improve the performance of x-ray free electron lasers,[5-9] ultrafast electron diffraction systems,[10-15] and dynamic transmission electron microscopes[16-18] – scientific instruments at the forefront of high space-time resolution dynamic structural studies of solid-state materials[19-21] and molecules.[22-25] For example, even a modest reduction of a pulsed electron beam's MTE at the entrance of an XFEL undulator by a factor of ~5 (emittance reduction of 2-3) is expected to increase its optical output beam brilliance or, alternatively, its emitted x-ray photon energy by an order of magnitude.[26,27] Similarly, for the case of electron diffraction, a lower MTE will improve the fidelity (i.e., quality) of the diffraction pattern due to the associated increase in the transverse coherence length of the electron beam.[28,29]

The brightness of an electron pulse containing $N$ electrons at emission from a photocathode is conventionally defined[30] as being proportional to $I/(\epsilon_{nx}\epsilon_{ny})$ with its peak current $I = Nq/\tau$: Here $Nq$ is the total pulse charge, $\epsilon_{nx}$ and $\epsilon_{ny}$ are the normalized transverse beam emittances (propagation in the $z$ direction), and $\tau$ is the rms electron pulse duration (essentially the time resulting from the convolution of the incident laser pulse shape generating the electron pulse and the temporal function associated with the 'response time' of the photocathode[31]). For an initial emitting rms transverse beam size $\sigma_x$ on the photocathode surface (e.g., the incident laser spot size) and circular beam symmetry, the one-dimensional ($x$) normalized transverse beam emittance can be expressed as $\epsilon_{nx} = \sigma_x\sqrt{\langle p_x^2\rangle}/(m_0 c)$,[32] where $\Delta p_x = \sqrt{\langle p_x^2\rangle}$ is the rms transverse momentum of the photoemitted electrons, $m_0$ is the free electron mass, and $c$ is the speed of light in vacuum. As the mean transverse energy of electrons emitted from the photocathode is defined by $MTE = \frac{1}{2m_0}\langle p_x^2 + p_y^2\rangle = \langle p_x^2\rangle/m_0$ for a circularly symmetric beam, the electron pulse's four-dimensional transverse phase space volume may be written as $\epsilon_{nx}\epsilon_{ny} = \sigma_x\sigma_y . MTE/(m_0c^2)$. Thus, since $N$ is given by photocathode's quantum efficiency (QE) multiplied by the number of absorbed photons in the incident laser pulse, the brightness of the electron pulse (beam) is proportional to the ratio $\frac{QE}{MTE}$. Although much prior effort has been put into the maximizing (and optimizing) the QE of photocathodes, with some semiconductor photocathodes in the 10% range,[33,34] reports of an MTE significantly below the 'thermal limit'[35] (e.g., 25 meV at 300 K [36,37] and 8 meV at 90 K [38]) have been scarce[39-41] despite theoretical predictions.[39,42-46]

In this paper, we present experimental results that indicate a pathway to sub-thermal electron emission; namely, the observation of a 6 meV MTE contribution to the total electron photoemission from a planar, iron-doped, single-crystal Gallium oxide photocathode at 300 K.

As indicated in prior studies on cesiated (i.e., activated) GaAs photocathodes,[39,40] such a low MTE signal is consistent with direct emission into the vacuum states from the lower conduction band (LCB) of $Ga_2O_3$ for which it is theoretically expected[46] that $MTE \approx \left(\frac{m_T^*}{m_0}\right) k_B T_e$ due to transverse momentum conservation[47]: Here $m_T^*$ is the transverse effective mass of the emitting band states, $k_B$ is Boltzmann's constant, and $T_e$ is the temperature of the emitting electron distribution determined by the cooling and transport dynamics in the photocathode. This 'inner' signal is superimposed on a stronger 'outer' signal with a larger MTE that is shown to be consistent with (optical)phonon-mediated Franck-Condon (FC) emission of thermalized electrons populating an upper conduction band (UCB) in $Ga_2O_3$ with a negative electron affinity $\chi$.[48,49] Methods of enhancing the quantum efficiency of the desired (ultra)low MTE inner signal with respect to the background FC emission are discussed.

Further, the presented spectral MTE measurements of the $Fe{:}Ga_2O_3(010)$ photocathode clearly display emission physics associated with two transport regimes.[48] Below an incident photon energy $\hbar\omega$ of 4.5 eV, where the sub-thermal 6 meV MTE inner signal is observed, the long transport regime is prevalent as photoexcitation into the LCB is from the optically active, near mid-gap Fe dopant states.[50] Above $\hbar\omega = 4.5$ eV, band-to-band absorption significantly reduces the absorption depth forcing the photocathode into the short transport regime in which both the inner and outer MTE signals are consistent with phonon-assisted FC emission. Moreover, in this latter regime, evidence is also presented that points to a rapid polaron establishment resulting in an increase in the electron temperature $T_e$ associated with the release of the polaron formation self-energy.[51,52] Consequently, in addition to providing insights into the energetic dynamics of polaron formation, the Fe-doped Gallium oxide photocathode provides an opportunity, in a single material, to study distinct photoemission mechanisms in two electron transport regimes characterized by different photoexcited carrier densities and dissimilar thermalized electron temperatures.

## Experimental and Theoretical Methods

The studied β-polymorph,[53] single-crystal, Gallium oxide sample was provided by Kyma Technologies Inc.[54] and is oriented for (010)-face emission. The ~7x7 mm sample is 450 μm thick and is obtained from a 2-inch diameter wafer preprepared for epitaxial growth (i.e., 'epi ready'). As a result, the emission face has a specified surface roughness Ra < 0.5 nm which ensures that photocathode surface roughness[55-57] does not affect the measurements to any significant extent.[48] The wide band gap $\beta$-$Ga_2O_3(010)$ semiconductor is also Fe doped to a level of $\sim 10^{18}$ $cm^{-3}$ to ensure semi-insulating behavior, generating two optically-active iron states at 3.05(±0.05) eV and 3.85(±0.05) eV below the primary LCB minimum.[50] The photocathode is therefore transparent in the visible with a weak absorption onset at ~400 nm due to transitions into the LCB from the upper Fe dopant state which pins the Fermi level.

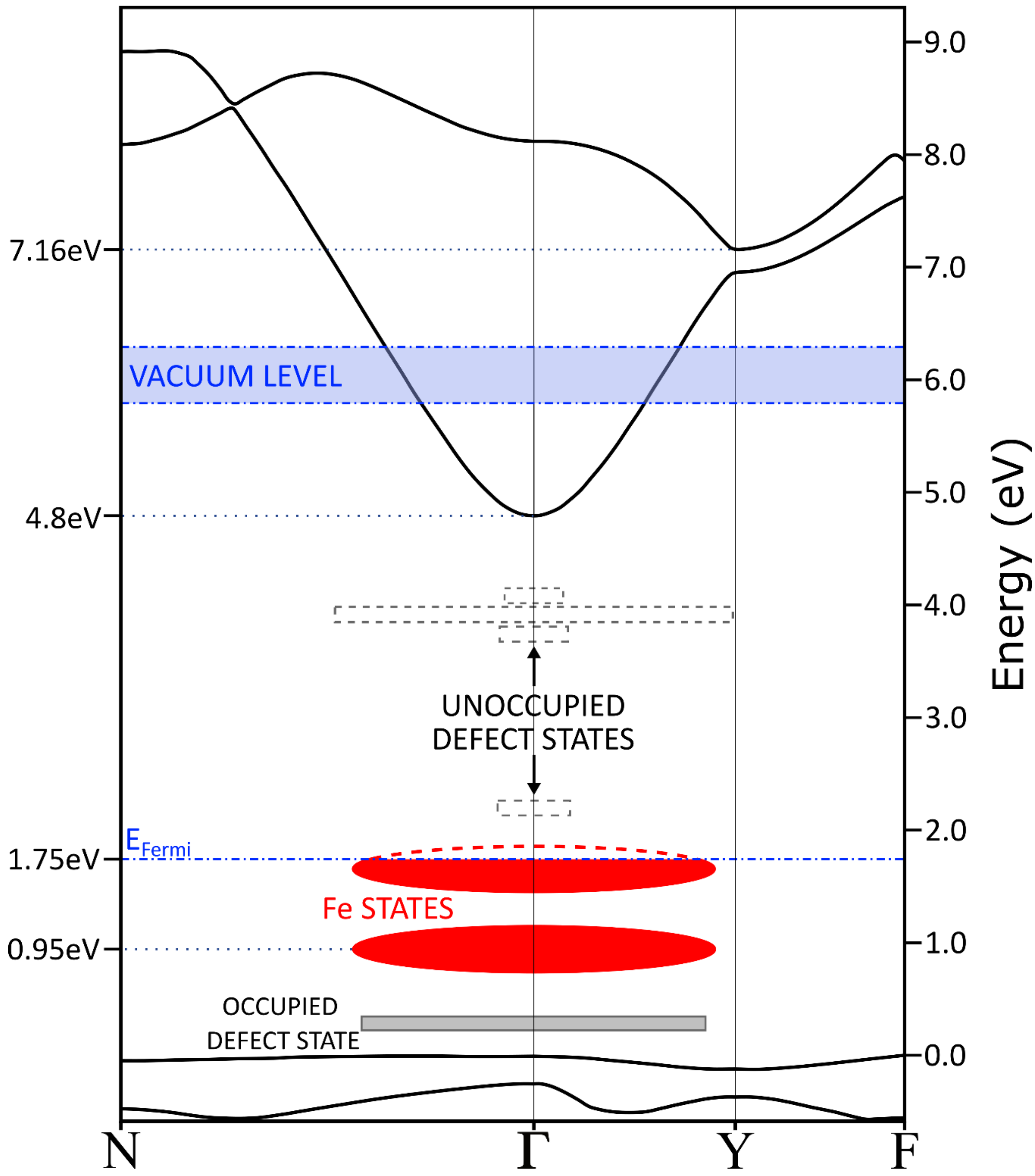

**Figure 1**

*Ab initio* calculated electronic band structure of Fe-doped β-$Ga_2O_3$ with zero energy at the valence band Γ point showing the measured energetic positions of the dopant states (red)[50] (with the upper Fe state pinning the Fermi level) and the positions of the occupied and unoccupied oxygen vacancy states.[58] The thin-slab calculated vacuum level at 6.0(±0.2) eV is depicted by the blue region and the important energies of the LCB and UCB minima are labeled.

Our in-house density functional theory (DFT) calculation of the electronic band structure of β-$Ga_2O_3$ is shown in Figure 1 together with the energetic positions of (i) the upper (pinning the Fermi level) and lower Fe dopant states and (ii) the occupied (below the Fermi level) and unoccupied known defect states (likely oxygen vacancies) of the oxide photocathode.[58] The *Ab initio* band structure evaluations are performed using the Amsterdam Modeling Suite's BAND module[59] with rSCAN-family functionals[60] from the LibXC library.[61] They indicate that the effective electron mass associated with the Γ-point dispersion of the LCB is $0.22m_0$, in agreement with literature values.[53] They also place the minimum of the next highest UCB at about 2.36 eV above the LCB minimum which itself is located at the 4.8 eV Γ-point band gap above the anisotropic valence band.[53] Together with thin slab calculations,[62] that place the vacuum level 6.0(±0.2) eV above the valance band (as shown in Figure1), this provides theoretical values of $\chi_{LCB}$ = 1.2(±0.2) eV and $\chi_{UCB}$ = −1.1(±0.2) eV for the electron affinities of the LCB and UCB respectively.

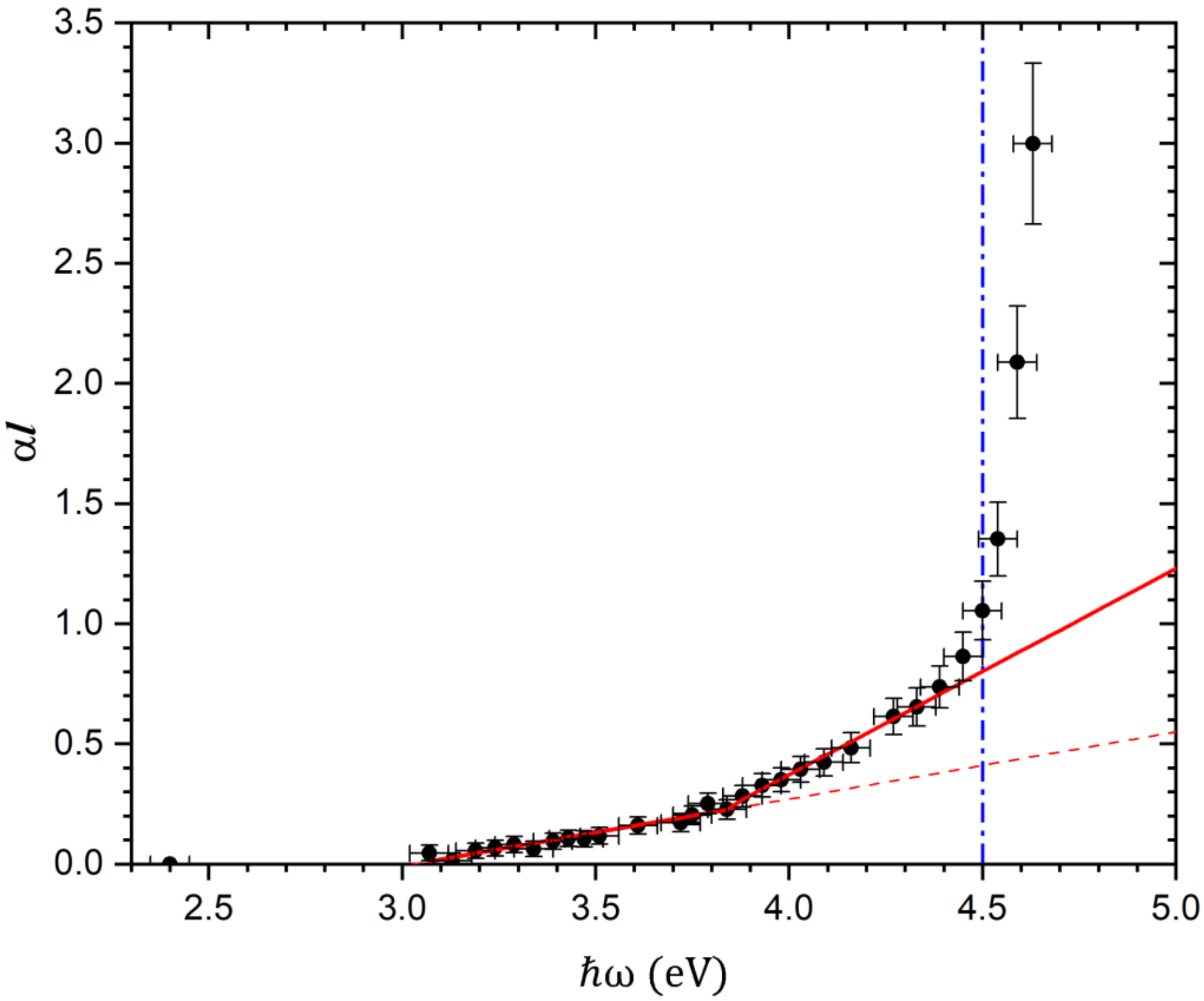

**Figure 2**
The absorption-length product, $\alpha l$, as a function of the photon energy ħω for the Fe-doped β-$Ga_2O_3$(010) photocathode determined from a transmission measurement.[50] Linear fits (red lines) below 4.5 eV to the $\alpha l$(ħω) data provide the 3.05 and 3.85 eV absorption thresholds for the two optically-active Fe dopant states. The vertical blue dot-dashed line at ħω = 4.5 eV indicates the onset of band-to-band absorption. [Reproduced from L.A. Angeloni, I.J. Shan, J.H. Leach, and W.A. Schroeder, "Iron dopant levels in β-$Ga_2O_3$," Appl. Phys. Lett. **124**, 252104 (2024), with the permission of AIP Publishing.]

The product of the absorption coefficient and the sample thickness, α*l*, for the studied Fe:$Ga_2O_3$(010) photocathode, evaluated as a function of the photon energy ħω using a transmission measurement,[50] is shown in Figure 2. The value of α*l* is less than unity for incident photon energies ħω below 4.5 eV and increases sharply for ħω > 4.5 eV. This leads to two electron transport regimes for the oxide photocathode.[48] For photon energies between the observed 3.05 eV 'threshold' for photoexcitation into the LCB from the upper Fe dopant state[50] and 4.5 eV, electrons are photoexcited into the LCB from the two Fe states over the entire 450 μm thickness of the photocathode, leading to a 'long' transport regime for the emitted electrons. Conversely, for ħω > 4.5 eV, the absorption depth becomes increasingly much shorter than the sample thickness, leading to a 'short' transport regime for the emitted electrons. The transmission measurement[50] yielding α*l* therefore also suggests that the 'effective' band gap of β-$Ga_2O_3$ could be as low as 4.5 eV due to both indirect absorption associated with the anisotropic valence band[53] and a probable significant Urbach tail, irrespective of the observed ~0.25 eV red-shift of the direct Γ-point band gap as the temperature is increased from near zero (the effective temperature of DFT band structure calculations) to the 300 K experimental temperature.[63,64] Based on the photon energy dependence of the direct band-to-band absorption for GaN,[65] which has a similar ~$0.2m_0$ effective mass for its LCB,[66,67] the absorption depth in this latter regime is expected to be quickly reduced to less ~100 nm just above ħω ≈ 4.5 eV.

These two transport regimes and the energetic positions (i.e., electron affinities) of the LCB and UCB with respect to the vacuum level are key to understanding the observed spectral dependence of the MTE for the Fe:$Ga_2O_3$(010) photocathode. Specifically, with a knowledge of the temperature $T_e$ of the electron distribution in the bulk photocathode, whereas direct one-step band emission predicts that $MTE \approx \left(\frac{m_T^*}{m_0}\right) k_B T_e$,[46] for momentum-resonant FC emission in the parabolic band approximation with the energy of phonon ħΩ → 0 one obtains

$$MTE = k_B T_e \left[\frac{3k_B T_e + \chi}{2k_B T_e + \chi}\right] \qquad (1)$$

for positive electron affinity (PEA) of the emitting band (χ > 0) and

$$MTE = \frac{|\chi|}{2} + k_B T_e \left[\frac{3k_B T_e + |\chi|}{2k_B T_e + |\chi|}\right] \qquad (2)$$

for negative electron affinity (NEA) (χ < 0).[48] For the case where the electron affinity is much greater that the energy of the photoexcited electron distribution ($|\chi| >> k_B T_e$), equations (2) and (3) reduce to $MTE \approx k_B T_e$ for PEA and $MTE \approx \frac{1}{2}|\chi| + k_B T_e$ for NEA. Further, for polar semiconductor photocathodes with strong optical deformation potential scattering,[68] as is the case for β-$Ga_2O_3$,[69] multi-phonon emission[49,70] is likely and a series solution evaluation for the MTE is generally required as outlined in Ref. 71.

The spectral dependence of the MTE for the Fe:$Ga_2O_3$(010) photocathode is measured at 300 K using a sub-picosecond tunable UV laser system coupled to a 10-20 kV DC gun-based photocathode characterization system as described in Refs. 72 and 73. Briefly, optical parametric amplification of a nonlinear fiber generated continuum, driven by a 20 W, 16.7 MHz femtosecond Yb:fiber laser system, is upconverted by sum frequency generation to produce ~0.5 ps tunable UV pulses. The generated *p*-polarized, 230-400 nm radiation with an average power of 10-100 μW is incident at 60° and focused to an irradiation area of $2\pi\sigma_x\sigma_y \approx 10^{-4}$ cm$^2$ (a near-Gaussian elliptical beam with rms widths of $\sigma_x \cong 50$ μm and $\sigma_y \cong 25$ μm) on the photocathode surface. The resultant incident peak pulse intensity of ~100 kW/cm$^2$ ensures that the observed photoemission is almost entirely due to one-photon absorption since the ~1 cm/GW two-photon absorption coefficient of β-$Ga_2O_3$ [74,75] implies that an intensity of greater than 1 MW/cm$^2$ is required for nonlinear absorption effects to approach the 1% level over the 450 μm sample thickness. The photoemitted electrons accelerated in the DC gun, based on the design by Togawa et al.,[76] enter a 42 cm drift region before detection using a 10 μm pore diameter dual microchannel plate/phosphor screen detector (BOS-18, Beam Imaging Solutions). The beam image on the P-43 phosphor screen is relay imaged with an 8:5 optical demagnification onto a 2.4 μm pixel CMOS digital camera (FL-20BW, Axoim Optics). The photocathode characterization system is calibrated for MTE measurements using accurate electron trajectory simulations[72] that are in very good agreement with analytical Gaussian (AG) electron beam propagation simulations of the experimental system.[77,78] For the ~25 μm point spread function (PSF) of the phosphor (~2 keV incident electrons), these propagation analyses predict a DC gun voltage dependent MTE resolution of less than 0.5 meV; specifically, at the employed DC gun voltage of 16 kV, the electron trajectory and AG models predict a transverse momentum calibration of 0.89 $(m_0.eV)^{1/2}$/mm on the CMOS camera and consequently a one-dimensional transverse electron energy resolution of better than 0.4 meV for the 25 μm phosphor PSF plus an additional smaller ~5 μm optical resolution of the achromat-based *f* /2 optical relay imaging system.

Before insertion into the DC electron gun, the surface of the Fe-doped β-$Ga_2O_3$(010) photocathode sample is cleaned using standard optical techniques (propan-2-ol and lens tissue) to remove physical surface contaminants and well as organic oils etc. After insertion into the DC electron gun and prior to the characterization measurements, the photocathode undergoes additional 'laser cleaning' for typically 20-30 minutes using the fourth harmonic radiation at 257 nm ($\hbar\omega$ = 4.82 eV) generated from the output of the sub-picosecond Yb-doped fiber laser. The effectiveness of the latter cleaning using ~1 W/cm$^2$ of UV-C radiation (to remove chemical absorbates, conceivably including hydrogen) is monitored by detecting the QE of electron emission and terminating the process once the QE has increased to a stable value. As oxidation effects can be neglected for a hard crystalline oxide like β-$Ga_2O_3$, this cleaning procedure is expected to produce a clean (perhaps reconstructed) native surface.

**Long Transport Regime**

Figure 3 displays the detected spatial electron beam profile for ħω = 4.33 eV which is in the long transport regime ($\alpha l < 1$, see Figure 2) for the Fe:$Ga_2O_3$ photocathode[48] (i.e., ħω < 4.5 eV). The horizontal section through the diameter (and peak) of the beam profile (right panel) is well described by a double Gaussian fit, giving MTE values of around 280 meV for the stronger 'outer' signal (blue Gaussian) and 6 meV for the weaker 'inner' signal (red Gaussian). The relative strengths and MTE values of these two signals were determined to be independent of the DC gun voltage, incident laser power and beam position on the photocathode, position of the electron beam on the MCP detector, MCP amplification voltages, and digital camera exposure time. We therefore conclude that the two detected emission signals correspond to two independent emission processes associated with the photoexcited electron dynamics (i.e., thermalization and transport) in the Fe-doped β-$Ga_2O_3$(010) photocathode. Under such circumstances, the total MTE of the emitted electron beam may be written as

$$MTE_{tot.} = \frac{MTE_1 . QE_1 + MTE_2 . QE_2}{QE_1 + QE_2} , \qquad (3)$$

where $QE_i$ and $MTE_i$ ($i$ = 1,2) are the quantum efficiencies and MTEs of the two emission processes. In this case (Figure 3), it is clear that $MTE_{tot.}$ is well approximated by the MTE of the stronger outer signal since $QE_{outer} >> QE_{inner}$ and $MTE_{outer} >> MTE_{inner}$. This is in fact the case over the entire investigated 3.4 to 5.3 eV photon energy range; that is, $MTE_{tot.} \cong MTE_{outer}$ in both the long and short transport regimes. It is also evident from Figure 3 that the detected electron beam exhibits a small ellipticity; 10-15% for the inner signal and 5-10% for the outer signal. The cause for this unknown but it is not expected to be associated with either (i) the 2:1 (major to minor axis ratio) elliptical shape of the tunable UV laser beam incident on the photocathode surface or (ii) aberrations in electron optics encountered by the photoemitted electron beam as it propagates through our experimental system. A simple analysis, which ignores the divergent electrostatic lensing of the anode in our DC gun, reveals that the elliptical photoelectron source with a 5 meV MTE generates a beam ellipticity of less than 2% on the MCP detector 42 cm from the DC gun. This is consistent with the detailed particle tracking analysis we have performed for the experimental system. Further, two co-propagating electron beams form the same photocathode source area should be affected in the same manner by the aberrations of the electrostatic anode lens (the only electron optic) in our system, yielding the same detected beam ellipticities. This is evidently not the case since the inner and outer beam ellipticities are observed to differ by roughly a factor of 2. We also note here that the described experimental system has been used to characterize the emission properties of numerous other photocathode materials (e.g., single crystal metals, semiconductors, bulk metallic glass, etc.), all generating near circular detected electron beams (less than 5% ellipticity) with the same experimental procedure; in particular, a metallized diamond(001) photocathode where electrons photo-injected from the back surface metal coating propagate over its 0.5 mm thickness before emission,[71] implying that even a recessed electron source at the back of an isotropic cubic crystalline photocathode does not produce an elliptical electron beam. Thus, we conjecture that the observed beam ellipticity may be related to the inherent electron emission physics of the bi-axial β-$Ga_2O_3$ crystal; for example, the difference in the dielectric constants in the crystal plane perpendicular to the (010) emission direction,[79] coupled perhaps to internal and surface fields generated by a photo-induced non-equilibrium charge redistribution associated with the high density of defect and dopant states in the crystal. Accordingly, in this paper, we report MTE values with uncertainties that include this beam ellipticity; for example, at $\hbar\omega$ = 4.33 eV (Figure 3), 6(±1) meV for the inner signal and 280(±25) meV for the outer signal.

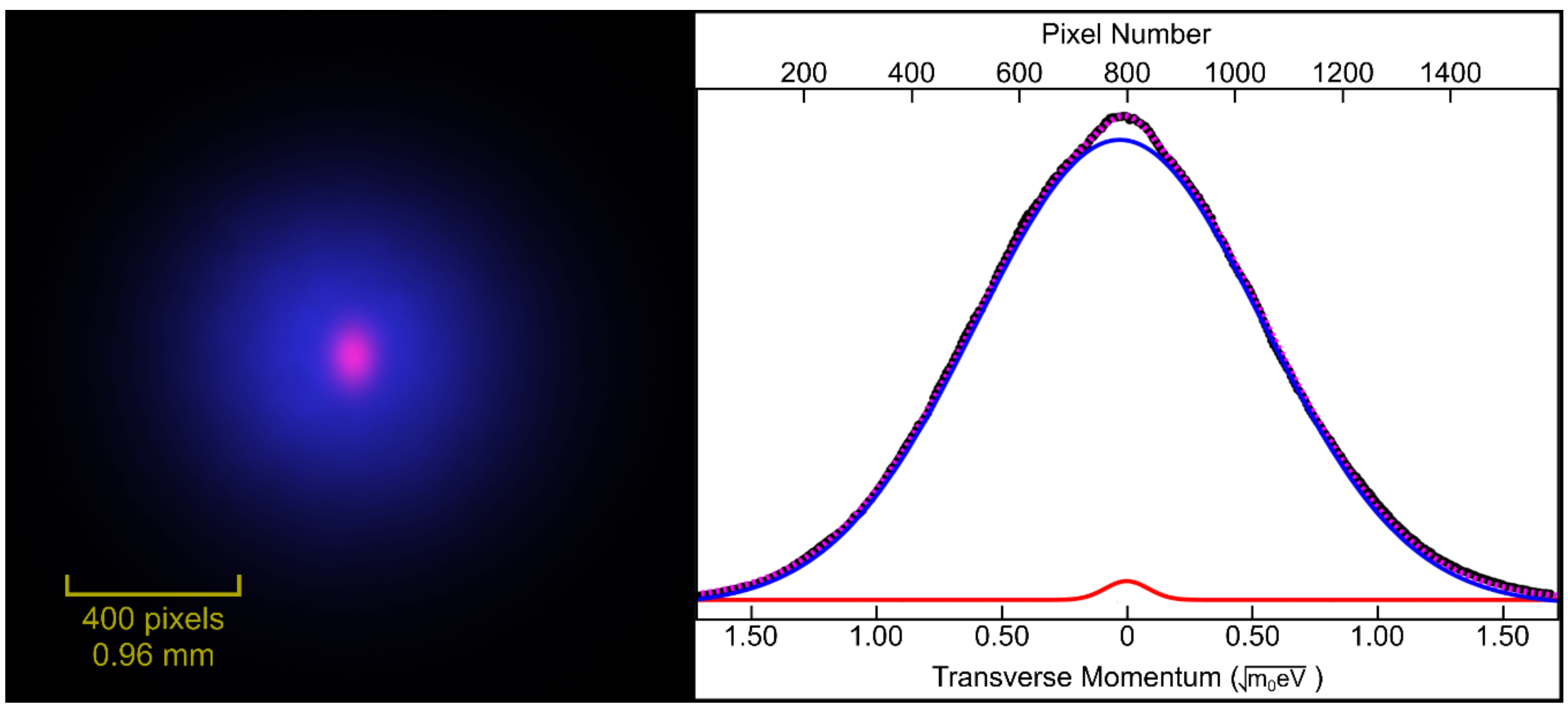

**Figure 3**
Two-component Gaussian signal decomposition.[71] Left panel: False color image of the raw background subtracted digital signal obtained at $\hbar\omega$ = 4.33 eV and for a 16 kV DC gun voltage showing the lower MTE inner signal (red) and higher MTE outer signal (blue). Right panel: Fitted Gaussian spatial beam profiles to a horizontal section through the peak of the digital data (black). The sum of the weaker inner (red) and stronger outer (blue) Gaussians generate the fit (purple dots). The ~7:1 ratio of the outer to inner beam widths immediately give a ~50:1 MTE ratio as the top horizontal pixel axis is directly related to the bottom transverse momentum axis; specifically, 0.00214 $(m_0.eV)^{1/2}$/pixel. From the two Gaussian fits with rms widths of 36(±6) and 247(±22) pixels respectively, the extracted inner signal MTE = 6(±1) meV and outer signal MTE = 280(±25) meV, where the uncertainties primarily reflect the beam ellipticity. The Gaussian fits therefore imply a ~500:1 ratio in the outer to inner signal strengths; that is, for this incident photon energy, the smaller inner signal is ~0.2% of the total signal.

The weaker but distinctly 'sub-thermal' (MTE < 25 meV) inner signal component emitted from the 300 K Fe:$Ga_2O_3$(010) photocathode is consistent with direct (band to vacuum) emission from the Boltzmann tail of the thermalized electron distribution in the LCB that is above the vacuum level (see Figure 1). Two factors lead to this conclusion. First, for the quoted ~300 $cm^2/(V.s)$ limiting electron mobility in $Ga_2O_3$ [69] and our applied 1.6 kV/cm internal field (16 kV/cm DC gun acceleration field and a dielectric constant[76] of around 10), the electron drift velocity to the emission face $v_d \leq 5{,}000$ m/s; that is, ~5% of the theoretical saturation drift velocity under optical deformation potential scattering[68] $v_{ds} \approx \sqrt{3\hbar\Omega/(4m^* \coth(\hbar\Omega/2T_L))}$ = 9.5×10$^4$ m/s for an electron in the LCB ($m^* \approx 0.22m_0$ [53]) at the $T_L$ = 300 K lattice temperature predominantly scattering with the strongest lowest energy ħΩ ≈ 25 meV optical phonon mode.[63,80-82] Based on electron transport analysis under these conditions[68] $T_e/T_L \cong 1 + v_d/v_{ds}$, so that the energy of the drifting electron distribution $k_BT_e \approx 27$ meV. Second, our DFT evaluations of the band structure of $Ga_2O_3$ reveal that the LCB in the vicinity of the vacuum level (1.2 eV above its minimum (Figure 1)) has a dispersion transverse to the (010) emission direction associated with an effective mass $m_T^*$ = 0.24(±0.02)$m_0$. For direct (band to vacuum) emission, this low transverse effective mass is predicted to restrict the transverse electron momentum states that can emit into the vacuum,[47] giving an expected constant $MTE \approx \left(\frac{m_T^*}{m_0}\right) k_BT_e$ of 6-7 meV [46] – clearly consistent with the experimental measurement.

Armed with the knowledge that $k_BT_e \approx 27$ meV in the long transport regime, the origin of the larger MTE outer signal in Figure 3 can also now be determined. First, we note that this signal cannot be due to (optical)phonon-mediated Franck-Condon electron emission from electrons above the vacuum level in the LCB. This is because for $\chi_{LCB} \approx +1.2$ eV and $k_BT_e \approx 27$ meV, equation (1) for PEA FC emission predicts an MTE of about 27 meV; that is, one order of magnitude less than that measured. Consequently, for a highly polar material like $Ga_2O_3$, the outer signal is very likely due to FC emission from the next highest band (Figure 1) – the UCB with a negative electron affinity (NEA) $\chi_{UCB} \approx -1.1$ eV. Indeed, the Fröhlich coupling constant of $g$ = 1.2 for the LCB of $Ga_2O_3$ [69] suggests that this is the case since the effective mass dependence of optical deformation potential scattering,[68] $g = \sqrt{m^*}g_0$ indicates a large value of about 2.5 for the 'intrinsic' ($m^* = m_0$) Fröhlich coupling constant $g_0$. In fact, for $k_BT_e \approx 27$ meV and $\chi_{UCB} \approx -1.1$ eV, our expanded multi-phonon ($g > 1$) NEA FC emission series summation analysis described in Ref. 71 can be used to fit the measured 280(±25) meV outer signal MTE using $g \approx 3.4$, assuming that the highest energy optical phonon with $\hbar\Omega_{max.} \approx 92$ meV [63,80-82] dominates the phonon-mediated emission process (i.e., has the strongest optical deformation potential scattering[68]). This theoretical interpretation reduces the magnitude of the UCB's NEA by $g\hbar\Omega_{max.} \approx 310$ meV, the real part of the self-energy shift associated with polaron formation for electrons in the UCB,[49,51,52] so that direct ($n$ = 0) emission from the UCB polaron has an effective electron affinity of −0.8 eV and the (optical)phonon-mediated FC emission is terminated by the lack of recipient vacuum states at $n$ = 8 phonon emissions.[71] (We note that the ±0.2 eV uncertainty our thin-slab-based[62] evaluation of the position of the vacuum level (Figure 1) generates a larger uncertainty in the value of $g$ for the UCB. For example, if $\chi_{UCB} \approx -0.9$ eV, the 280(±25) meV outer signal MTE value can be attained from the expanded ($g > 1$) multi-phonon FC emission analysis using $g \approx 2.2$ – a 35% lower value.)

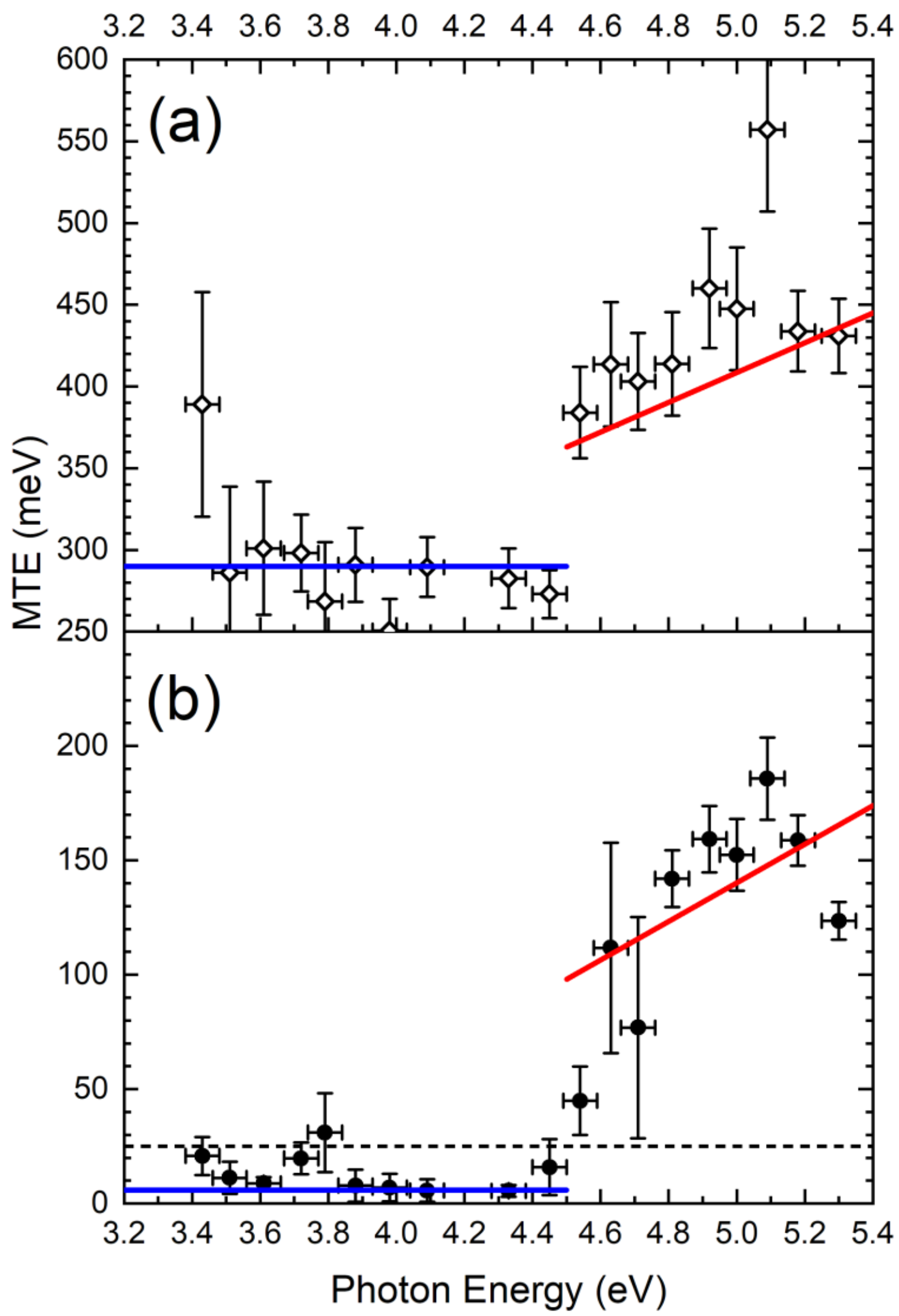

**Figure 4**
Measured spectral dependences of the MTE of the outer (a) and inner (b) signals resulting from the two-component Gaussian signal decomposition. Theoretical MTE dependences based on consideration of the electron temperature $T_e$ for the long and short transport regimes are shown by the blue and red lines respectively (see text). Also included is a dashed line at 25 meV for the thermal emittance associated with 300 K.

As shown by the blue lines in Figure 4, constant MTE values of 6 meV and 290 meV, which are consistent with the $k_B T_e \approx 27$ meV theoretical expectations for the inner (Figure 3(b)) and outer (Figure 3(a)) signals respectively, are good fits to the spectral MTE trend in the long transport regime below $\hbar\omega = 4.5$ eV. Small exceptions occur at (i) the lowest measured photon energies where the uncertainty in the extracted MTE values increases due to a reduced emission quantum efficiency and consequently lower signal-to-noise ratio, and interestingly, (ii) for the inner signal, around ħω ≈ 3.8 eV where photoexcitation into the LCB from the lower Fe dopant level becomes possible.[50] It is notable that in both cases a low population density of cold electrons is photoexcited into the bottom of the LCB.

The significantly ($10^2$-$10^3$×) higher QE of electron emission for the UCB is also expected, even though this band is weakly populated by a thermalized electron distribution with $k_B T_e \approx 27$ meV. First, for FC emission, its NEA nature provides for an intrinsic QE enhancement of around $10^{19}$ compared to emission from the LCB[48]; that is, an effective enhancement of perhaps $10^{17}$-$10^{18}$ as direct (band to vacuum) emission from the LCB dominates. Second, it has a larger density of states than the LCB leading to a relative population increase. And very importantly, third, the strong optical deformation potential scattering in β-$Ga_2O_3$ will rapidly replenish photoemitted electrons in the UCB, allowing for its much stronger FC electron emission rate to continue as electrons drift into the ~10 nm surface emission region.[48] This latter effect is further reinforced (especially in the short transport regime) by the energy equipartition between electron and optical phonon modes which results in a significant increase in the phonon population over that present at a 300 K lattice temperature.

**Short Transport Regime**

The spectral MTE data (Figure 4) also clearly exhibits the expected discontinuity at ħω = 4.5 eV separating the long and short electron transport regimes of the Fe:$Ga_2O_3$(010) photocathode. Above ħω = 4.5 eV in the short transport regime, the MTE of both the inner and outer signals increases linearly with photon energy after an abrupt step-like increase at the transition. Attention to the dependence of the electron energy $k_B T_e$ on incident photon energy ħω in this regime[48] along with consideration of the dominant emission mechanism allows for a consistent explanation for both signals.

The outer larger MTE signal (Figure 4(a)) displays a 70 meV increase at ħω = 4.5 eV indicating a rise in $k_B T_e$ to about 95 meV, which corresponds ~~well~~ to the highest energy $\hbar\Omega_{max.} \approx 92$ meV optical phonon in β-$Ga_2O_3$.[63,80-82] In fact, the red line in Figure 3(a) is obtained using the expanded multi-phonon analysis outlined above for FC emission[71] directly from the UCB states with $k_B T_e = \hbar\Omega_{max.} + \frac{2}{27}\left(\hbar\omega - E_g\right)$ and a band gap energy $E_g$ = 4.5 eV. This expression for $k_B T_e$ is consistent with that used in Ref. 48 for the short transport regime, where (i) a very rapid initial energy equipartition between the photoexcited electrons and optical phonons is presumed and (ii) the net optical phonon population decay time is assumed to be of the order of or less than mean time for electron emission. As the QE for NEA FC emission is proportional to at least $(k_B T_e)^2$,[48] the latter condition is strongly reinforced by the dominance of hot carrier emission, which will then be reflected in the measured MTE. Since β-$Ga_2O_3$ has $l$ = 8 dominant active optical phonon modes[63,80-82] and a relatively flat valence band dispersion (Figure 1),[53] the initial energy equipartition of the excess band-to-band photoexcited electron energy, ħω − $E_g$, yields the pre-factor of $\frac{2}{3(l+1)} = \frac{2}{27}$.[48] The additional electron energy contribution of $\hbar\Omega_{max.}$ = 92 meV to fit the experimental data is consistent with the formation of polarons in the LCB after photoexcitation – a step that is required for the drift transport of electrons to the emission face of the photocathode. In doing so, each electron transfers $g_{LCB}\hbar\Omega_{.}$ (the polaron formation self-energy) to the photoexcited 'carrier-polaron' distribution – resulting in a commensurate increase in $k_B T_e$ by energy conservation. In other words, although the photoexcitation is band-to-band, the experimental data supports the interpretation that following rapid energy equipartition with the optical phonon modes in this strongly polar semiconductor the photoexcited electrons form polarons for drift transport to the emission face; that is, occupy the polaron quasi-particle state[51,52] about 100 meV below the LCB. Our use of $\hbar\Omega_{max.}$ = 92 meV for the resultant released energy contribution is quite consistent with a weighted average optical phonon energy of 81(±2) meV evaluated from first-principles calculations for optical deformation potential scattering in β-$Ga_2O_3$ [82] multiplied by $g_{LCB}$ = 1.2.[69] (We note that this contribution to $k_B T_e$ is not observable in the long transport regime ($\alpha l < 1$) because the vast majority of electrons in the polaron quasi-particle would have time to cool to $k_B T_e \approx 27$ meV before emission.)

On the other hand, the inner signal MTE (Figure 4(b)) does not show a commensurate step-like increase at the ħω ≈ 4.5 eV transition between the long and short transport regimes: the increase is greater, from 6 meV to near 100 meV, instead of the expected $\left(\frac{m_T^*}{m_0}\right)\hbar\Omega_{max.} \approx 25$ meV for direct photoemission into the vacuum states.[46] This means that the emission process for the inner signal has changed across the 4.5 eV transition. Indeed, the new spectral MTE trend in the short transport regime is now consistent with FC emission from the large PEA LCB with the same photon energy dependence for $k_B T_e$ employed for the outer signal originating from the NEA UCB, yielding $MTE \approx \hbar\Omega_{max.} + \frac{2}{27}\left(\hbar\omega - E_g\right)$ as displayed by the red line in Figure 4(b). This change in emission process is readily explained by considering the LCB electron population distribution in each transport regime; specifically, the ratio in populations between $\chi_{LCB}$ and $\chi_{LCB} + \hbar\Omega_{max.}$ that cannot emit by FC emission using 92 meV optical phonons (as no vacuum states exist) and that at higher energies between $\chi_{LCB} + \hbar\Omega_{max.}$ and the UCB minimum at 1.1eV that can emit by FC emission from the LCB states. In the long transport regime with $k_B T_e \approx 27$ meV, 40 times more electrons reside in the lower energy bracket, giving a strong bias for direct emission from the LCB states. However, in the short transport regime with $k_B T_e \geq \hbar\Omega_{max.}$, this ratio is reduced to less than 2, implying that the emission mechanism with the larger transmission probability over the photoemission barrier will dominate; in this case, FC emission as its momentum resonance ensures unity transmission probability.[48]

Above ħω ≈ 5 eV, there is also some evidence in Figure 4 for a deviation of the MTE from the $\frac{2}{27}\left(\hbar\omega - E_g\right)$ linear trend, particularly for the inner signal. This is likely due to the photoexcitation of electrons into the LCB from a lower valence band located ~0.25 eV below the uppermost valence band at the Γ point of the Brillouin zone (Figure 1). As these electrons have a lower excess photoexcitation energy in the LCB, their presence will moderate the increase in $k_B T_e$ with ħω. However, since the density of states for the lower valence band is less than that of the upper valence band, the impact of this effect should be delayed until sufficient 'colder' electrons can be photoexcited – evidently until $\hbar\omega - E_g > 0.5$ eV.

**Discussion**

The performance of the studied Fe-doped β-$Ga_2O_3$(010) photocathode, including possible improvements associated with crystal growth advances and surface preparations to reduce χ, can be compared with that of other photoelectron sources with a view to, for example, application in ultrafast electron diffraction/microscopy (UED/M) instruments that require bright sources. First we note that when considered in isolation, the intrinsic beam brightness of the inner signal obtained from the Fe:$Ga_2O_3$(010) photocathode in the long transport regime at 300 K is similar to that observed for photoemission from a Cu(001) photocathode at 35 K;[83] the latter having an MTE of 5 meV and a QE ~ $10^{-8}$ while the 6 meV MTE inner signal from the oxide photocathode is associated with ~1% (or less) of the measured ~$10^{-6}$ total beam QE.[50] Further, unlike the Cu(001) photocathode, its strength relative to that of the outer signal could be enhanced by using surface treatments to reduce the oxide's work function (i.e., the LCB electron affinity $\chi_{LCB}$). A $\Delta\chi = -1$ to $-1.5$ eV reduction, by perhaps relatively robust surface methylation or hydrogenation,[49,84] would generate $\chi \approx 0$ for the LCB which should allow significantly more efficient direct emission of the Γ-point photoexcited electron population. This could theoretically produce a QE above 0.1% for the inner signal at $\hbar\omega \approx 4.4$ eV if electron recombination effects are not too deleterious in this long transport limit – as can certainly be expected as β-$Ga_2O_3$ crystal growth quality improves. This estimated QE increase needs to be verified experimentally, but is based on a factor of $\exp[|\Delta\chi|/(k_B T_e)] > 10^{16}$ increase in the emitting electron population above the vacuum level (for $k_B T_e \approx 27$ meV in the long transport limit) and prior measurements on cesiated GaAs that indicate a QE of around 1% for $\chi \approx 0$ under 808nm irradiation[41] (also the long transport limit[48]) – and that for a photocathode material with a ~6× lower LCB density of states than β-$Ga_2O_3$. In contrast, the QE of the outer larger MTE signal from the UCB should now be in the $10^{-5}$ range as $k_B T_e \approx 27$ meV,[48] thus ensuring that the ultra-low MTE inner signal could dominate by more than a factor of 100; the net result being a transverse momentum distribution (i.e., far-field beam profile) similar to that observed by Liu et al.[40] To ensure that optical phonon mediated FC emission from the LCB is still not present to a significant extent, a slightly positive value for $\chi_{LCB}$ of ~0.1 eV may need to be employed so that the electron population density between $\chi_{LCB}$ and $\chi_{LCB} + \hbar\Omega_{max.}$ still greatly exceeds that above $\chi_{LCB} + \hbar\Omega_{max.}$. If this cannot be achieved, FC emission from the LCB will contribute a signal with an MTE of about $\frac{3}{2}k_B T_e \approx 40$ meV for $\chi \approx 0$ (equations (1) and (2))[48] in the long transport regime, resulting in a net MTE of perhaps 20 meV. We note that a similar increase in the QE using surface treatments is also possible for Cu(001)[85,86] even at cryogenic temperatures, but at the expense of a significant increase in the MTE since for metals an increase in QE is only possible with an increase in the excess energy of photoemission.[32,35,87]

If the smaller MTE inner signal can be significantly enhanced with respect to the outer signal, then the issue of its separation from the latter will need to be addressed for practical applications. Fortunately, the required transverse momentum (or beam) filtering has already been successfully employed in UED/M systems[88-90] and to suppress higher transverse momentum dark current from a high-gradient RF gun.[91] The opportunity may therefore exist to substantially improve the performance of UED/M instruments using a surface-treated Fe-doped β-$Ga_2O_3$ photocathode with $\chi \approx +0.1$ eV. For example, such a filtered sub-10 meV electron beam would increase the spatial coherence of the electron beam by about a factor of 2 over that possible with metal photocathodes emitting near the photoemission threshold with a thermal limited MTE of near 25 meV[36,92] at 300 K and offer a greatly enhanced brightness. As the brightness of an electron beam is proportional to $\frac{QE}{MTE}$, even the brightness of the entire beam emitted in the long transport regime by the studied Fe:$Ga_2O_3$(010) photocathode with $\chi_{LCB} \approx +1.2$ eV (MTE ≈ 290 meV and QE ~ $10^{-6}$) is only an order of magnitude less than that of metal photocathodes at the same emitted MTE (e.g., QE ~ $10^{-5}$ for Rh(110)[72]). Moreover, expected future improvements in β-$Ga_2O_3$ crystal growth should reduce the defect density (e.g., oxygen vacancies (Figure 1)) and hence the photoexcited electron recombination rate, thereby increasing the QE of a Gallium oxide photocathode, especially in the long transport limit.

The performance of the investigated Fe:$Ga_2O_3$(010) photocathode, and its potential performance when $\chi \approx +0.1$ eV, also compares favorably with that of alkali antimonide photocathodes. Specifically, for $Cs_3Sb$ at 300 K, a minimum MTE value of 30 meV with a QE ~ $10^{-7}$ at $\lambda \approx 825$ nm has been reported,[93] which is consistent with a measured ~40 meV MTE at $\lambda = 690$ nm.[38] At these lowest MTE values, $Cs_3Sb$ therefore exhibits a QE to MTE ratio of the same order as that of the whole beam emitted by the studied Fe-doped β-$Ga_2O_3$(010) photocathode in the long transport limit. At shorter wavelengths, such as in the green ($\lambda \approx 520$ nm), the QE of $Cs_3Sb$ increases to ~1% while the MTE remains below 150 meV,[93,94] implying a ~$10^4$ brightness increase which is equivalent to that expected for Fe:$Ga_2O_3$(010) when its electron affinity is reduced to 0.1-0.2 eV. However, although demonstrating high brightness operation at visible and near infrared wavelengths, alkali antimonide photocathodes (i) do not offer the possibility of the described sub-thermal MTE performance (with appropriate beam filtering) through surface treatment and (ii) are expected to be less chemically robust than either a methylated or hydrogenated $Ga_2O_3$(010) surface.[49,84]

Further, the performance of other robust wide band gap photocathodes may be compared to that of the studied Fe:$Ga_2O_3$(010) photocathode. Ultrananocrystalline diamond photocathodes have a reported intrinsic emittance of around 1 μm/mm at 300 K under 262 nm ($\hbar\omega = 4.73$ eV) irradiation,[95] which corresponds to an MTE of about 500 meV, and a minimum measured MTE of 75 meV at 300 K under irradiation with ~4.25 eV photons ($\lambda \sim 290$ nm).[96] These measurements indicate that these photocathodes have an MTE comparable to, but generally greater than, the whole beam for our $Fe{:}Ga_2O_3(010)$ photocathode. Further, the photon energies required for electron emission are greater for the ultrananocrystalline diamond photocathode. Similarly, the total beam MTE of a single-crystal metallized diamond(001) photocathode[71] has recently been reported to be ~300 meV with a QE of around $10^{-6}$ for $\hbar\omega > 5$ eV giving a performance comparable to that of the Fe-doped $\beta$-$Ga_2O_3$ photocathode in the long transport limit. A *p*-type GaN(0001) photocathode without surface treatments also has a similar performance at 300 K,[48] even though it operates in the short transport limit.

Nanoscale plasmon-based ultrafast electron emitters[97] also have MTE values comparable to that of the entire beam emitted from the Fe-doped $\beta$-$Ga_2O_3(010)$ photocathode. Under one-electron-per-pulse emission, their reported normalized transverse emittance $\epsilon_{nx} = 40$ pm.rad which, for the 50 nm rms source size ($\sigma_x$), corresponds to an MTE of around 300 meV. Thus, the only reason for the small emittance value of the plasmon-based emitter is its sub-100 nm size. For the planar $Fe{:}Ga_2O_3(010)$ photocathode, a projected strong (and filtered) sub-10 meV MTE emission when $\chi \approx +0.2$ eV under diffraction-limited (laser) irradiation ($\sigma_x \approx 0.5\mu m$ in the near UV (250-300nm)) could produce $\epsilon_{nx} < 100$ pm.rad, which is comparable to the reported plasmon-based emitter. Moreover, under an extraction field of 10MV/m with $\sigma_x \approx 0.5\mu m$, Child's Law enforces a limit of about 100 electrons/pulse which would then correspond to a maximum 4D-brightness of ~$10^4$ electrons/($nm^2$Sr) – about 100× better than the maximum 4D-brightness reported for the HiRES UED beam line.[90,91]

The time response of the $Fe{:}Ga_2O_3(010)$ photocathode under our experimental conditions can also be estimated. Prior work on fitting the measured spectral dependence of the QE for this photocathode[50] when $\hbar\omega \leq 4.5$ eV indicates that the product of the electron drift velocity and recombination time, $v_d\tau$, is around 30 μm – the effective electron 'escape depth' in the long transport regime. For $v_d \leq 5{,}000$ m/s in our internal 1.6kV/cm applied field, this suggests a response time of ~10 ns. In an RF photoelectron gun, the higher field will lead to an electron drift velocity approaching the saturation drift velocity of ~$10^5$ m/s, reducing the response time to under 1 ns. However, such high-field transport will also result in an increase in the electron energy to around 1.5× the 738 K Debye temperature[63,68,98] giving $k_BT_e \approx 90$ meV so that direct band emission from the LCB should have an MTE of about 20 meV for $\chi \geq 0$. Improvements in $\beta$-$Ga_2O_3$ crystal growth will likely reduce the electron recombination rate into unoccupied oxygen vacancy states above the Fermi level (Figure 1) but thereby increase the response time. The use of an optically resonant half-wave thickness,[99] $\lambda/2n \approx 70$ nm [100] at $\hbar\omega \approx 4.4$ eV, for a higher-quality oxide crystal would cap any response time increase to ~1 ns while also improving the QE by a factor of perhaps 2-3 through an enhancement of the UV irradiance in the bulk photocathode material. Such an engineered $Ga_2O_3$ photocathode placed in a DC gun with a robust surface treatment to reduce $\chi_{LCB}$ by 1 eV to ~0.1 eV could then be expected to yield an electron beam with a sub-10 meV MTE at 300 K and with a QE of ~0.1%.

**Summary**

The measured spectral photoemission properties of a single-crystal and Fe-doped β-$Ga_2O_3$(010) photocathode display a diversity of emission mechanisms that are well described by theoretical expectations. Foremost, the experimental investigations at 300 K have revealed a small ultracold 6 meV MTE contribution to its photoemission in the long transport regime ($\alpha l < 1$) occurring for $\hbar\omega \leq 4.5$ eV. This contribution is due to direct emission from the LCB and consistent with theoretical expectations[46] indicating that $MTE \approx \left(\frac{m^*}{m_0}\right) k_B T_e$ with an electron energy $k_B T_e \approx 27$ meV reflective of the expected drift transport.[68] Although superimposed on a ~~~10×~~ stronger background emission with an MTE of ~290 meV, when considered in isolation, the observed inner signal has a transverse beam brightness comparable to that measured near threshold for a Cu(001) photocathode at 35 K.[83] Its QE efficiency relative to the larger signal background, due to optical phonon mediated FC emission[48] from a NEA UCB with $\chi_{UCB} \approx -1.1$ eV and $k_B T_e \approx 27$ meV, could be significantly improved by using surface treatments to reduce the work function (i.e., $\chi_{LCB}$) by ~1 eV – potentially resulting in sub-10 meV MTE emission with a QE ~ 0.1%.

For incident photon energies ħω greater than 4.5 eV, direct band-to-band absorption results in a transition to the short transport regime ($\alpha l >> 1$), where the higher density of photoexcited electrons equipartition their excess band energy with the $l = 8$ dominant optical phonon modes in β-$Ga_2O_3$,[48,63,80-82] resulting in the MTEs of both the inner and outer signal being consistent with FC emission using $k_B T_e = \hbar\Omega_{max.} + \frac{2}{27}\left(\hbar\omega - E_g\right)$ and a band gap energy $E_g = 4.5$ eV. The transition into the short transport regime therefore also resulted in the inner signal emission converting from direct band emission to phonon-mediated FC emission – a shift that is readily explained by a change in electron population distribution to higher band energies as a result of the step-like increase in $k_B T_e$ that is consistent with the release of the self-energy $g_{LCB}\hbar\Omega \approx \hbar\Omega_{max.}$ associated with polaron formation. Nonetheless, the inner signal MTE remains below ~150 meV for $\hbar\omega < 5.3$ eV.

In addition to exhibiting a small ultracold (sub-77 K) contribution to the total emitted electron beam at 300 K, the single-crystal Fe:$Ga_2O_3$(010) photocathode also demonstrated a clear transition between the long and short electron transport regimes in a single photoemitting material: the higher electron density short transport regime being associated with band-to-band photoexcitation, while observation of the low electron density long transport regime was made possible by the ~$10^{18}$ $cm^{-3}$ iron doping necessary to achieve semi-insulating electrical behavior. A similar spectral transition between transport regimes should also be observed in cesiated GaAs photocathodes[39-43] where, due to the low LCB density of states ($m^* = 0.067 m_0$ [101]), the absorption depth for 808 nm light is ~1 μm (long transport regime[48]) and decreases to less than 100 nm below 500 nm.[101] Thus, the presented experimental data together with its self-consistent analysis should be invaluable to Monte-Carlo methods simulating photoemission from GaAs-based photocathodes[45] by providing a bridging description between direct band and Franck-Condon emission. On the other hand, semiconductor photocathodes such as GaN[48] (and β-$Ga_2O_3$) with a direct band gap and a LCB electron effective mass greater than about $0.2m_0$ are likely to exhibit short transport regime behavior as soon as direct band-to-band electron photoexcitation is possible since their absorption depth is rapidly reduced below 100 nm for $\hbar\omega > E_g$.

## Acknowledgements

This work was supported by the U.S. Department of Energy under Award no. DE-SC0020387.

## Author Declarations

*Conflict of Interest*

The authors have no conflicts to disclose.

*Author Contributions*

**Louis. A. Angeloni:** Conceptualization (equal); Data curation (lead); Formal analysis (lead); Investigation (equal); Software (lead); Visualization (lead); Writing – original draft (supporting); Writing – review & editing (equal). **I-J. Shan:** Conceptualization (supporting); Data curation (supporting); Formal analysis (supporting); Writing – review & editing (supporting). **J. H. Leach:** Resources (lead); Writing – review & editing (equal). **W. Andreas Schroeder:** Conceptualization (equal); Funding acquisition (lead); Project administration (lead); Supervision (lead); Writing – original draft (lead).

## Data Availability

The experimental data supporting the findings of this study are openly available in https://doi.org/10.25417/uic.29319281.v1, reference number 102, and the employed theoretical FC emission analysis is published.[48,71]